\documentclass[aps,twocolumn]{revtex4-1}
\usepackage{amsmath}
\usepackage{amssymb}
\usepackage{graphicx}% Include figure files
\usepackage{bm}% bold math
\usepackage{textgreek}

\makeatletter

\DeclareMathOperator\artanh{artanh}

\makeatother

\begin{document}

\title{Electric field lines of relativistically moving point charges}

\author{Daja Ruhlandt, Steffen M\"{u}hle, and J\"{o}rg Enderlein}
\email{jenderl@gwdg.de}
\affiliation{Third Institute of Physics -- Biophysics, Georg August University, 37077 G\"{o}ttingen, Germany}

\date{\today}
\begin{abstract}
Generation of electromagnetic fields by moving charges is a fascinating topic where the tight connection between classical electrodynamics and special relativity becomes particularly apparent. One can gain direct insight into the fascinating structure of such fields by visualizing the electric field lines. However, the calculation of electric field lines for arbitrarily moving charges is far from trivial. Here, we derive an equation for the director that points from the retarded position of a moving charge towards a specific field line position, which allows for a simple construction of these lines. We analytically solve this equation for several special but important cases: for an arbitrary rectilinear motion, for the motion within the wiggler magnetic field of a free electron laser, and for the motion in a synchrotron.
\end{abstract}

% \pacs{01.55.+b}{General physics}
% \pacs{03.30.+p}{Special relativity}
% \pacs{03.50.De}{Classical electromagnetism, Maxwell equations}

\maketitle 

\section{Introduction}Electric and magnetic fields generated by arbitrarily moving point charges are a fascinating topic where relativistic physics meets classical electrodynamics. In particular, accelerated point charges are the generators for almost all electromagnetic radiation, such as that emitted by oscillating electric dipoles, synchrotrons, or free electron lasers. As is well known, the electric field $\mathbf{E}$ of an arbitrarily moving point charge $q$ can be found with the help of Li\'enard-Wiechert potentials and has the explicit form \cite{landau1971classical,jackson1971classical}

\begin{equation}
\begin{split}
    \mathbf{E}\left(\mathbf{r},t\right) &= q \left\{\rule{0cm}{0.8cm}\right. \frac{\mathbf{R}-R\boldsymbol{\beta}}{\gamma^2 \left(R-\mathbf{R}\cdot\boldsymbol{\beta}\right)^3} + \\  
    & \hspace{2cm} + \frac{\mathbf{R}\times\left[\left(\mathbf{R}-R\boldsymbol{\beta}\right)\times\dot{\boldsymbol{\beta}}\right]}{c \left(R-\mathbf{R}\cdot\boldsymbol{\beta}\right)^3} \left.\rule{0cm}{0.8cm}\right\}_{t'}    
\end{split}
\label{eq:Efield}
\end{equation}

\noindent where the three-dimensional vector $\mathbf{R}$ is the spatial part of the four-dimensional null-vector 

\begin{equation}
    \left\{ c (t - t'), \mathbf{r} - \mathbf{r}_0(t') \right\}.
    \label{eq:nullvector}
\end{equation}

\noindent This null-vector defines the retarded time $t'<t$ via

\begin{equation}
    t - t' = \frac{R}{c} = \frac{\left\vert \mathbf{r} - \mathbf{r}_0(t') \right\vert}{c}
    \label{eq:retardedtime}
\end{equation}

\noindent at which the right hand side of equation \eqref{eq:Efield} has to be evaluated. Here, $\mathbf{r}_0(t)$ is the particle's trajectory as a function of time $t$. Furthermore, the symbol $\boldsymbol{\beta}(t)=c^{-1}d\mathbf{r}_0(t)/dt$ is the particle's velocity divided by the speed of light $c$, $\gamma$ is the usual Lorentz factor $\gamma = 1/\sqrt{1-\beta^2}$, and a dot denotes differentiation after time. For finding the electric field at a given position $\mathbf{r}$ and time $t$, one has firstly to solve the retarded time equation \eqref{eq:retardedtime}, and then secondly to evaluate the right hand side of \eqref{eq:Efield} at time $t'$, which is typically a numerically demanding task. 

Another way of visualizing an electric field is to use electric field lines -- continuous lines tangential to the electric field vector. Visualization of field lines can help in better understanding complex field configurations generated by non-trivial particle trajectories, and knowledge of field lines can also be used to estimate the electric field strength, due to the interconnection between local field line density and field strength as embodied by the zero divergence of the electric field in source-free space. Thus, the question how to efficiently calculate and draw field lines for arbitrarily moving point charges has been repeatedly considered in the literature \cite{tsien1972pictures,ohanian1980electromagnetic, jefimenko1994direct,bolotovskiui1997force,singal2011first,datta2016b,franklin2019electromagnetic}.  Here, we present an efficient and relatively simple way of how to find and draw electric field lines of an arbitrarily moving charge by deriving a compact auxiliary equation for a unit vector pointing from the retarded position of the charge to a specific field line position. We then find analytic solutions of the problem for several important cases. 

\begin{figure}
\centering
\includegraphics[width=0.35\textwidth]{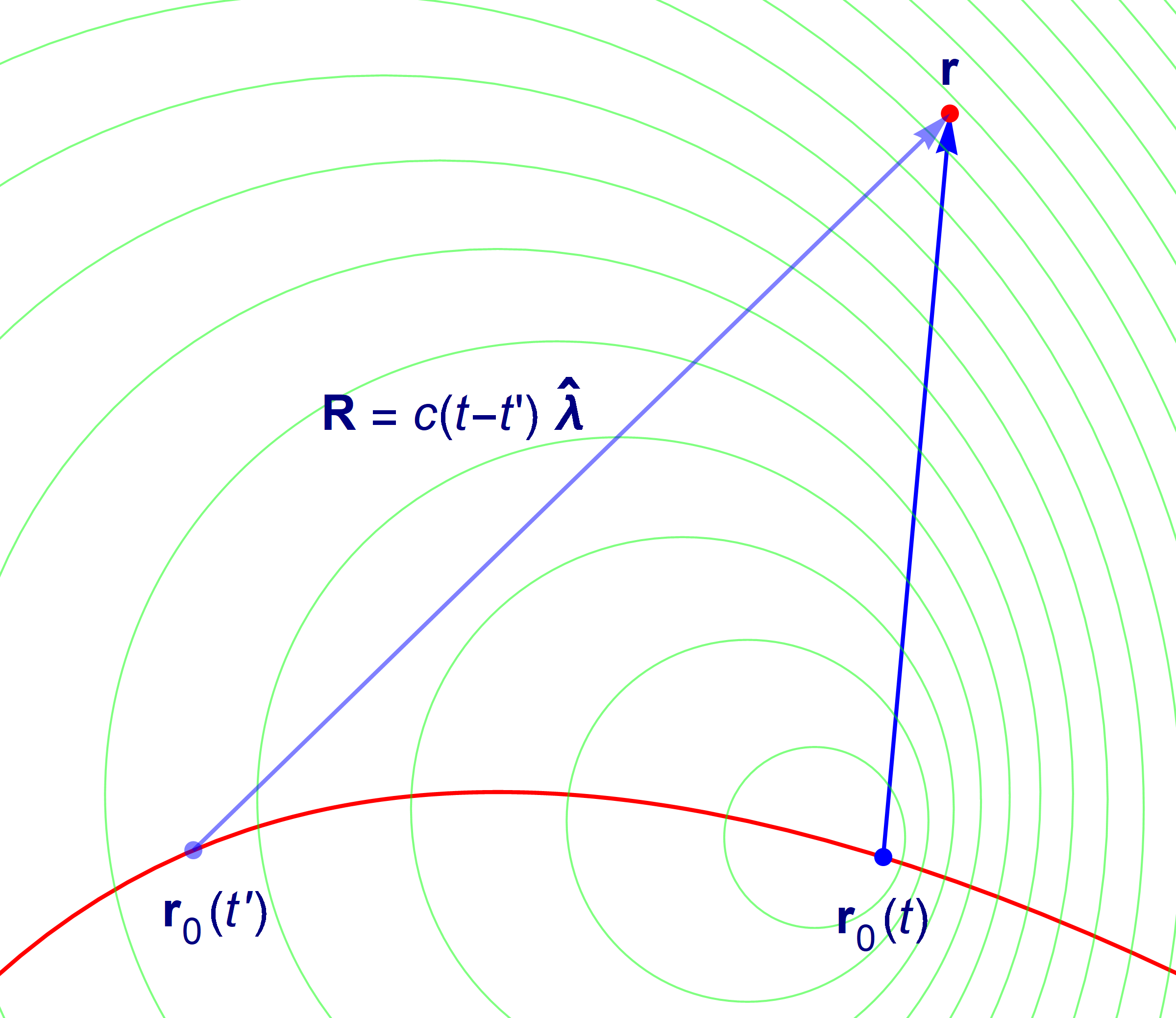}
\caption{A point charge moving along an arbitrary trajectory $\mathbf{r}_0(t)$ (red solid line) generates an electromagnetic field throughout space. The field at any given position $\mathbf{r}$ at time $t$ originates from the charge when it is at retarded position $\mathbf{r}_0(t')$. Green circles are lines of constant retarded time $t'$. The vector $\mathbf{R}=c(t-t')\hat{\boldsymbol{\lambda}}$ is the spatial part of the null-vector of eq.~\eqref{eq:nullvector} which connects the space-time events $\left\{c t',\mathbf{r}_0(t')\right\}$ and $\left\{c t,\mathbf{r}\right\}$, so that $\hat{\boldsymbol{\lambda}}$ is a unit vector.}
\label{fig:Fig1}
\end{figure}

\section{Arbitrary motion} Let us describe a field line at time $t$ by a parametric three-dimensional curve $\mathbf{p}\left(s\right)$ which is parametrized by the variable $s$. Along all its positions, it has to be parallel to the electric field vector, which means that it has to obey the differential equation 

\begin{equation}
    \frac{d\mathbf{p}\left(s\right)}{ds} \propto \mathbf{E}\left[\mathbf{p}\left(s\right)\right].
    \label{eq:fieldlineparametric}
\end{equation}

\noindent Taking into account the non-trivial form of the electric field as given by eq.~\eqref{eq:Efield}, finding analytic solutions to this equation seems to be a formidable task. Note that any Cartesian position $\mathbf{r}$ can geometrically be referenced to the retarded position $\mathbf{r}_0(t')$ by $\mathbf{r} = \mathbf{r}_0(t') + \mathbf{R}(t')$, where $t'$ is the retarded time of the particle's position when it contributes to the electric field at position $\mathbf{r}$, see also figure \ref{fig:Fig1}. In particular, this holds true also for positions $\mathbf{r}=\mathbf{p}(s)$ on a field line. Our core idea is to use the retarded time $t'$ to parametrize a field line, by setting $s=t'$. Thus, the time $t'$ has a double meaning: it denotes the retarded time $t'$ \emph{and} it parametrizes the field line, and we find for the field line positions the relation

\begin{equation}
    \mathbf{p}(t') = \mathbf{r}_0(t') + \mathbf{R}(t') = \mathbf{r}_0(t') + c(t-t') \hat{\boldsymbol{\lambda}}(t')
    \label{eq:pdef}
\end{equation}

\noindent where we have used the fact that the length of the vector $\mathbf{R}(t')$ is $c(t-t')$, so that the vector $\hat{\boldsymbol{\lambda}}(t')$ on the right hand side of eq.~\eqref{eq:pdef} is a unit vector pointing from the retarded position $\mathbf{r}_0(t')$ of the point charge to a position $\mathbf{p}(t')$. 
Now, let us consider eq.~\eqref{eq:fieldlineparametric}. Because we require that vector $d\mathbf{p}/ds\equiv d\mathbf{p}/dt'$ and vector $\mathbf{E}$ have only to be parallel at all positions $\mathbf{p}(t')$, we can choose any proportionality factor in eq.~\eqref{eq:fieldlineparametric} between these two vectors. Let us thus set $d\mathbf{p}/dt'$ equal to $-c\gamma^2 R^2 \left(1-\hat{\mathbf{R}}\cdot\boldsymbol{\beta}\right)^3 \mathbf{E}/q$ so that we find the field-line determining equation

\begin{equation}
    \frac{d\mathbf{p}\left(t'\right)}{dt'} = c \left(\hat{\mathbf{R}}-\boldsymbol{\beta}\right) + \gamma^2 \mathbf{R}\times\left[\left(\hat{\mathbf{R}}-\boldsymbol{\beta}\right)\times\dot{\boldsymbol{\beta}}\right]
    \label{eq:peq}
\end{equation}

\noindent where a hat over a vector symbolizes normalization (unit vector). Now, by inserting eq.~\eqref{eq:pdef} into the last equation, we find the following auxiliary equation for the unit vector $\hat{\boldsymbol{\lambda}}(t')=\hat{\mathbf{R}}$

\begin{equation}
    \frac{d\hat{\boldsymbol{\lambda}}}{dt'} = \gamma^2 \left[ \left( \hat{\boldsymbol{\lambda}} - \boldsymbol{\beta} \right) \times \dot{\boldsymbol{\beta}} \right] \times \hat{\boldsymbol{\lambda}}.
    \label{eq:lambda}
\end{equation}

\noindent This equation is the core result of our paper: When we can solve this equation and determine $\hat{\boldsymbol{\lambda}}(t')$ for all times $t'<t$, then we can use eq.~\eqref{eq:pdef} to find the full field line. Thus, $t'$ plays the role of a line parameter and does not have to be found \emph{a priori} from an implicit retarded time equation such as eq.~\eqref{eq:retardedtime}, as has to be done when calculating the electric field. The final condition of eq.~\eqref{eq:lambda}, i.e. the direction $\hat{\boldsymbol{\lambda}}(t'=t)$, defines into which direction a field line starts from a point charge at time $t$.

Although we cannot present a general solution of eq.~\eqref{eq:lambda} for an arbitrary motion $\mathbf{r}_0(t')$, we consider in the next chapters several important and quite general cases for which analytical solutions can be found.

\section{Rectilinear motion}
Let us assume that the velocity and acceleration are all the time co-linear, i.e. $\dot{\boldsymbol{\beta}}\parallel \boldsymbol{\beta}$. In that case, our auxiliary equation for $\hat{\boldsymbol{\lambda}}$ simplifies to 

\begin{equation}
    \frac{d\hat{\boldsymbol{\lambda}}}{dt'} = \gamma^2 \left( \hat{\boldsymbol{\lambda}} \times \dot{\boldsymbol{\beta}} \right) \times \hat{\boldsymbol{\lambda}} = \gamma^2 \left[ \dot{\boldsymbol{\beta}} - \hat{\boldsymbol{\lambda}} \left( \dot{\boldsymbol{\beta}} \cdot \hat{\boldsymbol{\lambda}} \right) \right] 
    \label{eq:lambdarecti}
\end{equation}

\noindent Multiplying both sides with unit vector $\hat{\boldsymbol{\beta}}$ leads to an equation for the component $\lambda_\parallel = \hat{\boldsymbol{\beta}}\cdot \hat{\boldsymbol{\lambda}}$ of $\hat{\boldsymbol{\lambda}}$ parallel to the constant direction of motion,

\begin{equation}
    d\lambda_\parallel = \frac{1 - \lambda_\parallel^2}{1-\beta^2}d\beta
    \label{eq:lambda_para}
\end{equation}

\noindent This equation can be integrated and has the solution

\begin{equation}
    \lambda_\parallel = \frac{n_\parallel + \beta}{1+n_\parallel \beta}
    \label{eq:lambda_para_sol}
\end{equation}

\noindent where $n_\parallel$ is an integration constant. For a $\hat{\boldsymbol{\lambda}}$-component $\lambda_\perp$ that is orthogonal to the direction of motion, we can find a similar equation by multiplying both sides of eq.~\eqref{eq:lambdarecti} with a unit vector perpendicular to $\hat{\boldsymbol{\beta}}$. This results in 

\begin{equation}
    d\lambda_\perp = - \frac{\lambda_\perp \lambda_\parallel}{1-\beta^2}d\beta = - \frac{\lambda_\perp \left(n_\parallel + \beta\right)}{\left(1+n_\parallel \beta\right)\left(1-\beta^2\right)}d\beta
    \label{eq:lambda_perp}
\end{equation}

\noindent which can also be explicitly integrated and has the solution

\begin{equation}
    \lambda_\perp = \frac{n_\perp}{\gamma\left(1+n_\parallel \beta\right)}
    \label{eq:lambda_perp_sol}
\end{equation}

\noindent with a second integration constant $n_\perp$. By adding $\lambda_\parallel^2$ and $\lambda_\perp^2$ together, one can check that $n_\parallel^2+n_\perp^2=1$ so that the integration constants are the components of a unit vector $\hat{\mathbf{n}}$. Putting all together, this leads to the compact result

\begin{equation}
\begin{split}
    \hat{\boldsymbol{\lambda}} = \frac{\hat{\mathbf{n}} + (\gamma-1) (\hat{\mathbf{n}}\cdot\hat{\boldsymbol{\beta}}) \hat{\boldsymbol{\beta}} + \gamma\boldsymbol{\beta} }{\gamma(1+\boldsymbol{\beta}\cdot\hat{\mathbf{n}})}
\end{split}
\label{eq:lambdares}
\end{equation}

\noindent Inserting this expression into eq.~\eqref{eq:pdef} and after some algebraic transformations, one finds the result for the electric field line itself as 

\begin{equation}
\begin{split}
    &\mathbf{p}\left(t'\right) = \mathbf{r}_0(t') + c(t-t') \boldsymbol{\beta}(t') +\\
    &\qquad + c(t-t')\left[\frac{(\gamma^{-1}-1)(\hat{\mathbf{n}}\cdot\hat{\boldsymbol{\beta}}) \hat{\boldsymbol{\beta}} + \hat{\mathbf{n}}}{\gamma\left(1+\hat{\mathbf{n}}\cdot\boldsymbol{\beta}\right)}\right]_{t'}
    \label{eq:final}    
\end{split}
\end{equation}

\noindent Here, $\boldsymbol{\beta}$ and $\gamma$ in the square bracket are evaluated at time $t'$. Please note that the expression $\mathbf{r}_0(t') + c(t-t') \boldsymbol{\beta}(t')$ in the above equation would represent the position of the moving charge \emph{if} it would continue to move uniformly with its instantaneous velocity $\dot{\mathbf{r}}_0(t')=c\boldsymbol{\beta}(t')$ from its position $\mathbf{r}_0(t')$ at time $t'$. Thus, the vector in the second line of eq.~\eqref{eq:final} points from this virtual position to the field line position corresponding to $t'$.   

The found expression for $\mathbf{p}\left(t'\right)$ gives an explicit parametric representation of a field line at time $t$, where the parametric variable is the retarded time $t'$. For finding a particular field line, one first defines $\hat{\mathbf{n}}$ and then traces the line for decreasing values of $t'$ starting from $t'=t$. 

To better understand the physical meaning of the unit vector $\hat{\mathbf{n}}$,
let us check eq.~\eqref{eq:final} against the well-known case of a point charge moving uniformly with velocity $\dot{\mathbf{r}}_0=c\boldsymbol{\beta}$. For this case, the electric field reads

\begin{equation}
\mathbf{E}(\mathbf{r},t) = \frac{q \gamma \Delta\mathbf{r}}{\left\{\gamma^2 (\Delta\mathbf{r}\cdot\hat{\boldsymbol{\beta}})^2 + \left[\Delta\mathbf{r}-(\Delta\mathbf{r}\cdot\hat{\boldsymbol{\beta}}) \hat{\boldsymbol{\beta}}\right]^2\right\}^{3/2}} 
\label{eq:efielduniform}
\end{equation}

\noindent where we have used the abbreviation $\Delta\mathbf{r} = \mathbf{r}-\mathbf{r}_0(t')-c(t-t')\boldsymbol{\beta}$. This expression describes an isotropic electric field which is "squeezed" by a factor $\gamma^{-1}$ along the direction of motion. Thus, if a field line is directed along unit vector $\hat{\mathbf{n}}$ in the particle's rest frame, it will point along direction 

\begin{equation}
\hat{\mathbf{n}}' = \frac{(\gamma^{-1}-1)(\hat{\mathbf{n}}\cdot\hat{\boldsymbol{\beta}}) \hat{\boldsymbol{\beta}} + \hat{\mathbf{n}}}{\sqrt{1-(\hat{\mathbf{n}}\cdot\boldsymbol{\beta})^2}}
\label{eq:unitvecuniform}
\end{equation}

\noindent in the observers' lab frame. Comparing eq.~\eqref{eq:unitvecuniform} with eq.~\eqref{eq:final} shows that eq.~\eqref{eq:final} indeed describes straight field lines along directions $\hat{\mathbf{n}}'$ starting from the instantaneous position $\mathbf{r}_0(t')+c (t-t')\boldsymbol{\beta}$ of the uniformly moving charge at time $t$, and that $\hat{\mathbf{n}}$ in eq.~\eqref{eq:lambda} is the starting direction of the field line within the rest frame of the moving charge. To summarize, eq.~\eqref{eq:final} describes the field line position as pointing from the virtual position $\mathbf{r}_0(t')+c(t-t')\boldsymbol{\beta}$ into the direction of the squeezed unit vector in the charge's rest frame. Thus, if $\beta=\text{const.}$, this direction is also constant and the field lines are straight lines originating from the virtual position at time $t$ of the charge.

In what follows, we consider several special case of rectilinear motion and calculate images of the corresponding field lines. All numerical calculations for the figures in this paper have been done with \textsc{Mathematica}, and the code can be found at Ref.~\cite{mmaprogram}. Animated GIFs for all the discussed examples below can be found at Ref.~\cite{movies}.

\subsection{Uniformly accelerated motion} As a first application of eq.~\eqref{eq:final} we consider the well-known classical example of a uniformly accelerated charge which is at rest at time zero, then (relativistically) accelerates along the (horizontal) $x$-direction with constant acceleration to the speed $c/\sqrt{2}$ within one unit of time, and then continues to move uniformly with that constant velocity. For such a motion, the particle's $x$-position as a function of time is given by 

\begin{equation}
    x_0(t) =\begin{cases} 0, & \mbox{if } t \leq 0 \\ c\left(\sqrt{1+t^2}-1\right),  & \mbox{if } 0<t\leq 1 \\ c\left(\sqrt{2}-1+(t-1)/\sqrt{2}\right),  & \mbox{if } t>1 \end{cases}
    \label{eq:accelerated}
\end{equation}

\begin{figure}
\centering
\includegraphics[width=6.5cm]{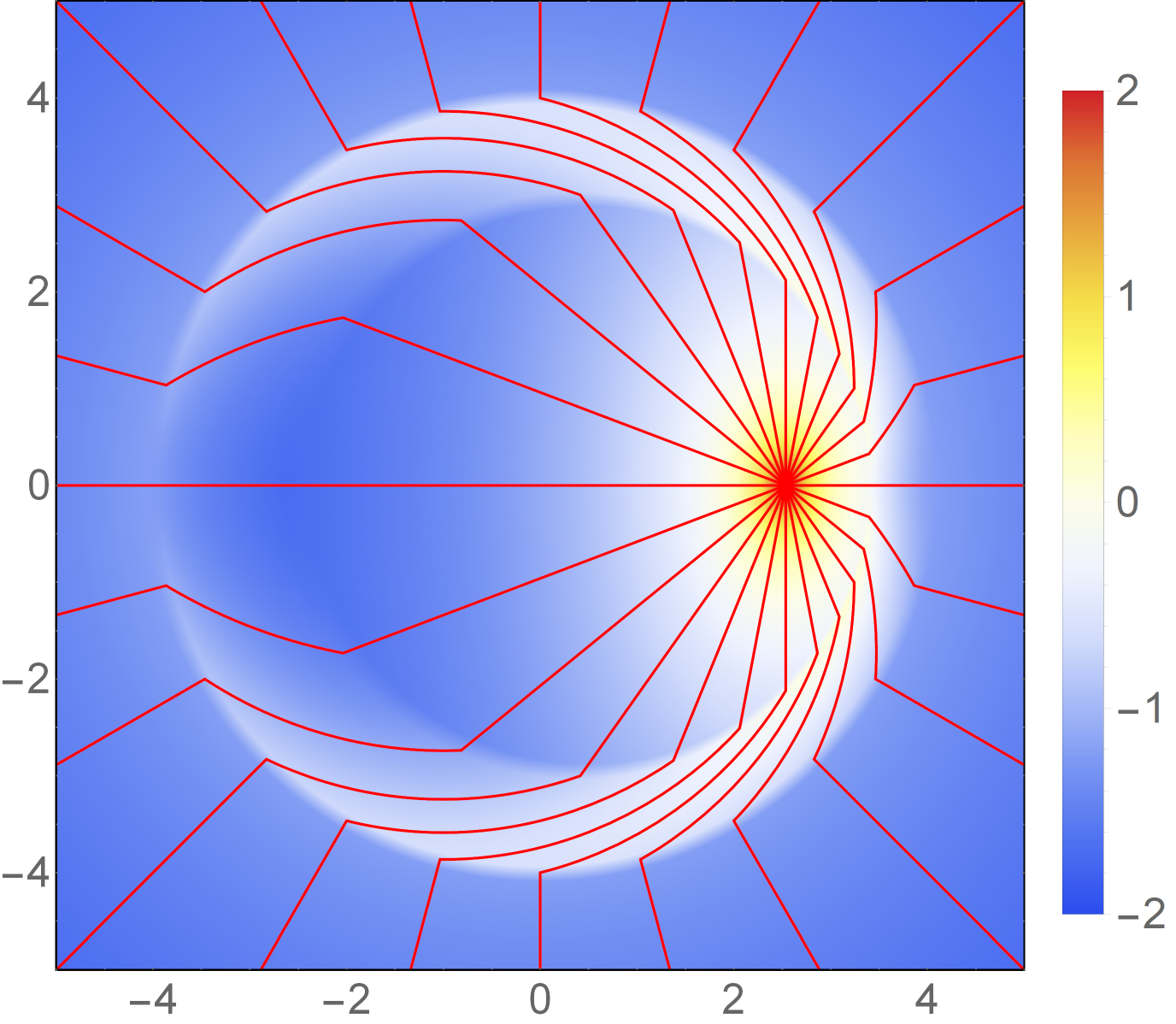}
\caption{Electric field (density plot) and electric field lines (red) for an accelerated point charge. Its position along the horizontal axis ($x$-axis) is given by eq.~\eqref{eq:accelerated}. The shown picture refers to time $t=4$, and the unit of length is chosen in such a way that the numerical value of the speed of light is one. The coloring encodes the decadic logarithm of the electric field amplitude $\vert\mathbf{E}\vert$ in arbitrary units. Here, we show field lines that start, in the particle's rest frame, from its position at angles $\phi=15^\circ$ to $\phi=360^\circ$ with respect to the horizontal axis in steps of $15^\circ$.}
\label{fig:Fig2}
\end{figure}

\noindent Fig.~\ref{fig:Fig2} shows the electric field lines overlaid with a density plot of the decadic logarithm of the electric field amplitude for the time $t=4$. As can be seen, eq.~\eqref{eq:final} nicely reproduces the field lines of the static charge at large distances and those of the uniformly moving charge at small distances from the particle, with the acceleration-related transition zone in between.  

\subsection{Bremsstrahlung} The second example considers the opposite situation: A uniformly moving charge (uniform speed $c/\sqrt{2}$) starts to decelerate at time zero with constant deceleration so that it stops moving at time one. Now, its position is given by

\begin{equation}
    x_0(t) =\begin{cases} c t/\sqrt{2}, & \mbox{if } t \leq 0 \\ c\left(1 +t/\sqrt{2} - \sqrt{1+t^2}\right),  & \mbox{if } 0<t\leq 1 \\ c\left(1+1/\sqrt{2}-\sqrt{2}\right),  & \mbox{if } t>1 \end{cases}
    \label{eq:decelerated}
\end{equation}

\begin{figure}
\centering
\includegraphics[width=6.5cm]{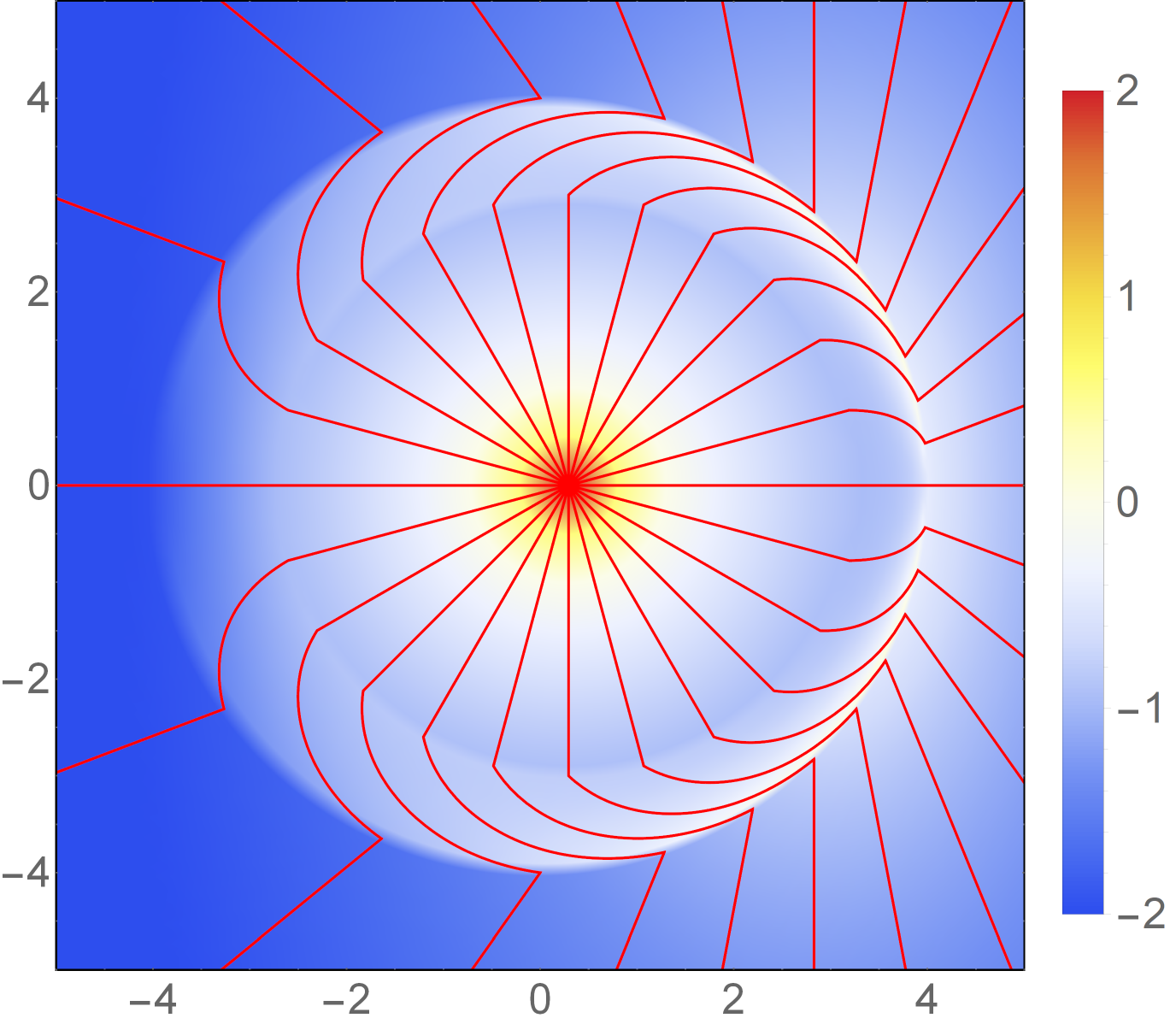}
\caption{Same as Fig.~\ref{fig:Fig2} but for a decelerating charge, the horizontal position of which is described by eq.~\eqref{eq:decelerated}. Shown is a snapshot for time $t=4$.}
\label{fig:Fig3}
\end{figure}

\noindent The resulting field lines and electric field for $t=4$ are presented in Fig.~\ref{fig:Fig3}. Although the motion of the charge is a simple time-reversal of the first example, the field lines and electric field look significantly different, which is, of course, a direct consequence of the retarded time effect. 

\subsection{Oscillating charge}

\begin{figure}
\centering
\includegraphics[width=6.5cm]{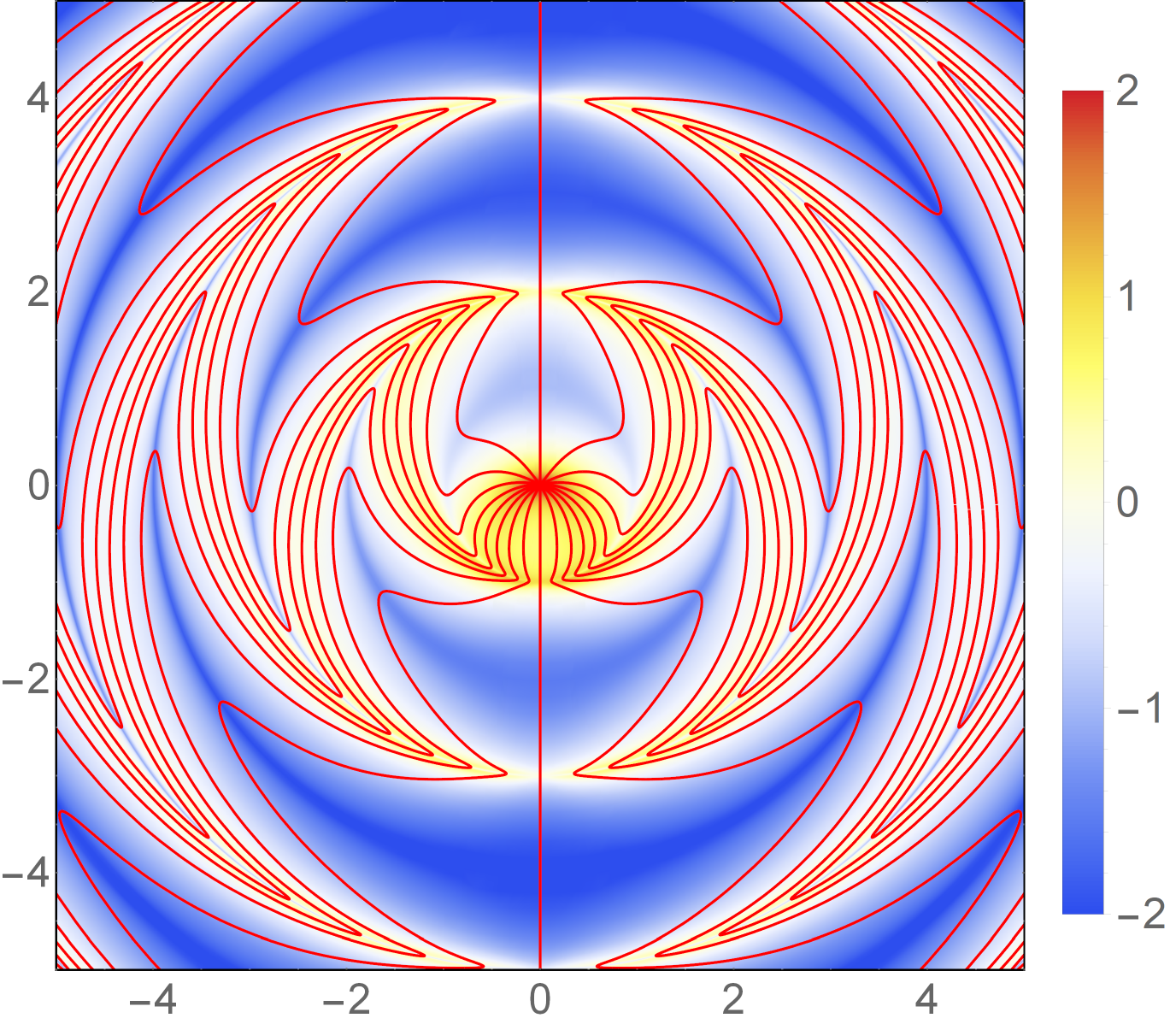}
\caption{Same as Fig.~\ref{fig:Fig2} but for an vertically oscillating motion, see eq.~\eqref{eq:oscillating}, with oscillation angular frequency $\omega=\pi$, and oscillation amplitude $a$ chosen in such a way that the maximum value of $\beta=\omega a/c$ is equal to 0.8. Shown is a snapshot at time $t=0$. Shown are field lines that start, in the particle's rest frame, from its position at angles $\phi=10^\circ$ to $\phi=350^\circ$ with respect to the horizontal axis in steps of $20^\circ$.}
\label{fig:Fig4}
\end{figure}

As a third and last example for rectilinear motion we consider a charge which oscillates with angular frequency $\omega$ and amplitude $a$ along the vertical direction. Thus, its vertical position is given by 

\begin{equation}
    y_0(t) = a \sin{\omega t}
    \label{eq:oscillating}
\end{equation}

\noindent so that $\hat{\boldsymbol{\beta}}=\hat{\mathbf{y}}$ and $\beta=a\omega\cos{\omega t}/c$. The resulting field lines and electric field for the numerical values $\omega=\pi$ and $\omega a/c = 0.8$ are presented in Fig.~\ref{fig:Fig4} for $t=0$. 

\section{Planar motion}

Remarkably, the expression of eq.~\eqref{eq:final} was already presented by Arutyunyan in the eighties of the last century, see refs.~\cite{arutyunyan1985force,arutyunyan1986electromagnetic,arutyunyan1992equivavlent}. There, it was claimed that it is valid also for non-rectilinear motions such as that of a charge in a synchrotron. Unfortunately, this is not true because its derivation was based on the assumption that $\boldsymbol{\beta}\parallel\dot{\boldsymbol{\beta}}$ so that the term with $\boldsymbol{\beta}\times\dot{\boldsymbol{\beta}}$ in eq.~\eqref{eq:lambda} drops out. In the next sections we consider two cases of planar motion (motion confined to a plane) where this is no longer true, but where we can still find analytical solutions to eq.~\eqref{eq:lambda}.

\subsection{Free electron laser}

Let us consider the motion of a point charge in a free electron laser~\cite{hopf1976classical,colson1977one,brau1996undulator}: A point charge moves with constant speed $c \beta_0$ along the $x$-direction and wiggles along the orthogonal $y$-direction with arbitrarily time-dependent velocity $c \beta_\perp(t)$. Thus, we now have 

\begin{equation}
    \boldsymbol{\beta}(t) = \beta_0 \hat{\mathbf{x}} + \beta_\perp(t) \hat{\mathbf{y}}
\end{equation}

\noindent and

\begin{equation}
    \dot{\boldsymbol{\beta}}(t) = \dot{\beta}_\perp(t) \hat{\mathbf{y}}
\end{equation}

\noindent In what follows, we consider always field lines in the plane of motion so that  $\hat{\boldsymbol{\lambda}}$ lies in the $xy$-plane. Let us denote the angle between $\hat{\boldsymbol{\lambda}}$ and the horizontal $x$-axis by $\psi(t)$ so that $\hat{\boldsymbol{\lambda}}\times\dot{\boldsymbol{\beta}} = \dot{\beta}_\perp \cos\psi(t) \hat{\mathbf{z}}$ where $\hat{\mathbf{z}}$ is a unit vector pointing out off the $xy$-plane. Also, we have $\boldsymbol{\beta}\times\dot{\boldsymbol{\beta}} = \dot{\beta_\perp}(t) \beta_0 \hat{\mathbf{z}}$. Taking into account that $d\psi/dt'$ equals the modulus of $d\hat{\boldsymbol{\lambda}}/dt'$ we find 

\begin{equation}
    \frac{d\psi}{dt'} = \frac{\dot{\beta}_\perp(t')}{1-\beta_0^2 - \beta_\perp^2(t')} \left( \cos\psi - \beta_0 \right) .
\end{equation}

\noindent This equation can be solved analytically and has the explicit solution 

\begin{equation}
\begin{split}
    \psi(t') &=\; 2\arctan\Bigg\{\frac{1}{\zeta} \tanh\bigg[\frac{1}{2} \artanh\left[\gamma_0\beta_\perp(t')\right]- \\
    &-\frac{1}{2} \artanh\left[\gamma_0\beta_\perp(t)\right] + \artanh\left(\zeta \tan\frac{\phi}{2}\right)\bigg] \Bigg\}
\end{split}
\end{equation}

\noindent where $\phi$ is the final angle of $\psi(t')$ at time $t'=t$, and where we have introduced the abbreviations 

\begin{equation}
\zeta=\sqrt{\frac{1+\beta_0}{1-\beta_0}}    \quad \text{and} \quad \gamma_0=\sqrt{1-\beta_0^2}.
\end{equation}

\noindent Knowing the solution for $\psi$, the unit vector $\hat{\boldsymbol{\lambda}}$ is given by

\begin{equation}
    \hat{\boldsymbol{\lambda}}(t') = \left\{\cos \psi(t'), \sin \psi(t') \right\}
\end{equation}

\noindent which, when inserted into eq.~\eqref{eq:pdef}, solves the problem of finding a parametric description for the electric field lines. The found expressions are valid for an arbitrary transverse motion described by $\beta_\perp(t)$. Let us consider the special case of a harmonic transverse oscillation with frequency $\omega$. Then, the particle's trajectory is described by

\begin{equation}
    \mathbf{r}_0(t) = c \beta_0 t \hat{\mathbf{x}} + \frac{c \beta_\perp}{\omega} \sin(\omega t) \hat{\mathbf{y}}.
    \label{eq:wiggling}
\end{equation}

\begin{figure}
\centering
\includegraphics[width=6.5cm]{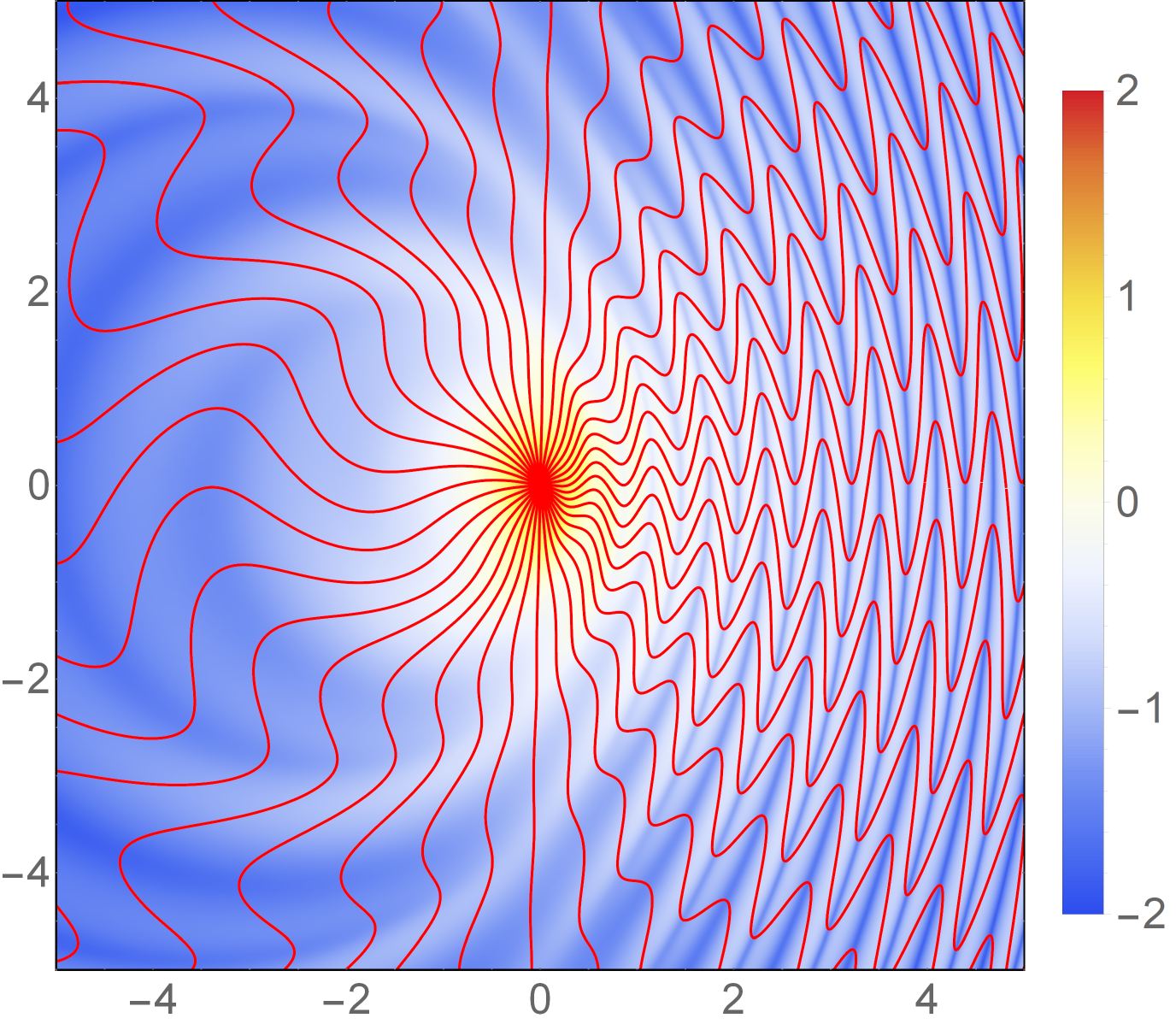}
\caption{Same as Fig.~\ref{fig:Fig2} but for a wiggling motion as described by eq.~\eqref{eq:wiggling}. Shown is a snapshot at time $t=0$ in the $xy$-plane.}
\label{fig:Fig5}
\end{figure}

\noindent Let us consider the following numerical values: $\omega=\pi$, $\beta_0 = 1/\sqrt{2}$ and $\beta_\perp = 0.1$. Thus, the particle moves uniformly with $1/\sqrt{2}$ light speed along the horizontal axis while oscillating with maximum 0.1 light speed vertically. The resulting field lines and electric field are shown in Fig.~\ref{fig:Fig5}. One nicely sees that regions of strong transverse field-line orientation (with respect to the line of sight from the particle) and thus field line density correspond to regions of large electric field strength.

\subsection{Synchrotron}

\begin{figure}
\centering
\includegraphics[width=6.5cm]{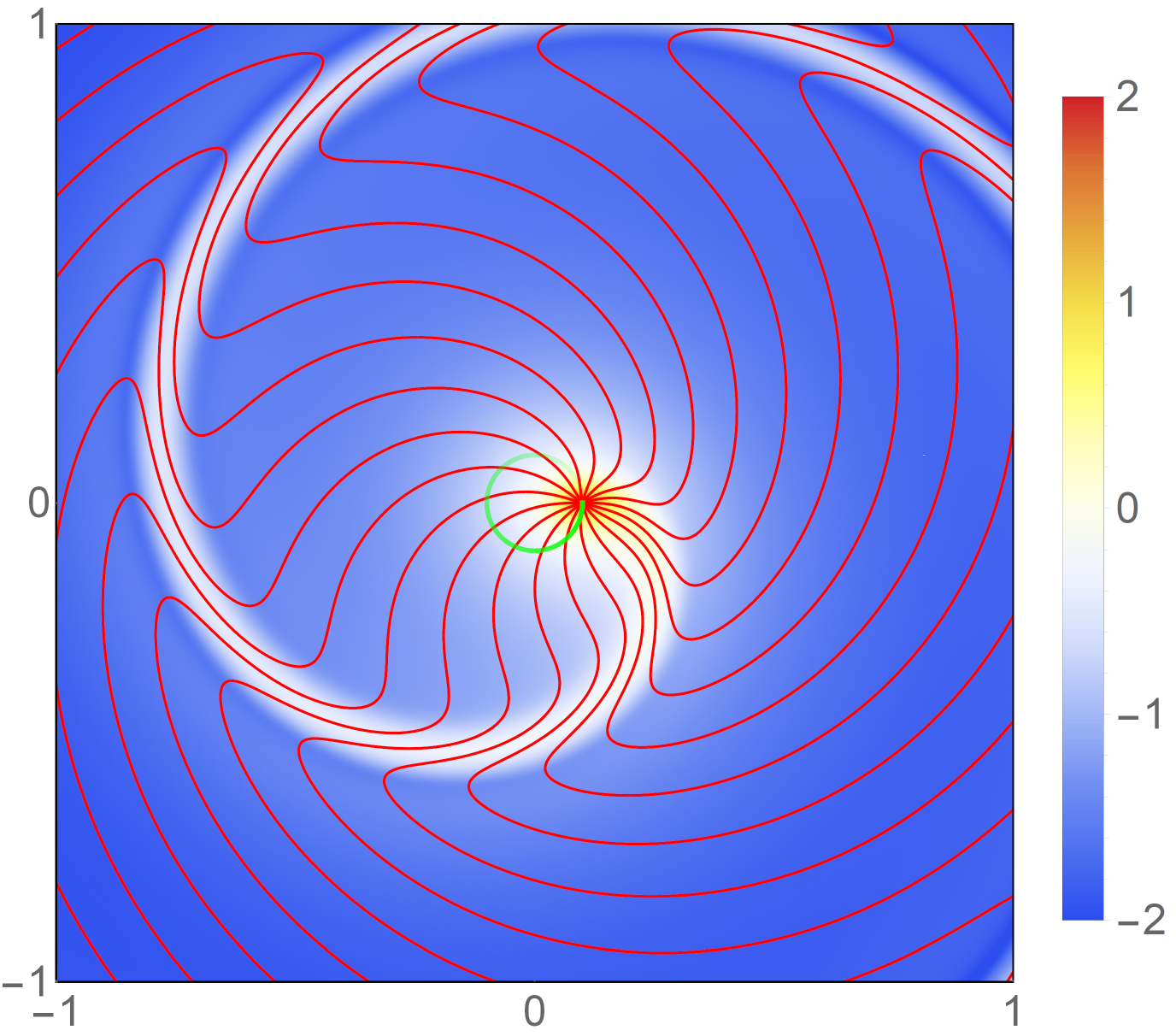}
\caption{Electric field lines and electric field amplitude for a circularly moving point charge (indicated by the green line) at 0.6 light speed. Here, the unit of length is 1~km, and the radius of the circular motion is 0.1~km. High field intensities coincide with strong bunching of electric field lines, demonstrating nicely the tight connection between field intensity and field line density.}
\label{fig:Fig6}
\end{figure}

The last considered example refers to the motion of a point charge in a synchrotron \cite{winick1978synchrotron,hofmann2004physics,hannay2005electric,novokhatski2011field,patterson2011simplified}: a motion with uniform speed around a circle with radius $a$ and angular frequency $\omega$. Thus, the time-dependent coordinate of the particle is described by

\begin{equation}
    \mathbf{r}_0(t) = a \left( \cos\omega t \; \hat{\mathbf{x}} + \sin\omega t \; \hat{\mathbf{y}} \right)
\end{equation}

\noindent so that the constant modulus of $\boldsymbol{\beta}$ is $\beta = a\omega/c$ and the constant modulus of $\dot{\boldsymbol{\beta}}$ is $\dot{\beta} = \omega\beta = a\omega^2/c$. Let us denote the angle between $\hat{\boldsymbol{\lambda}}$ and $\mathbf{r}_0$ by $\psi$. Then we find the determining equation for $\psi(t')$ from eq.~\eqref{eq:lambda} as

\begin{equation}
    \frac{d\psi}{dt'} + \omega = \gamma^2\left(\dot{\beta} \sin\psi -\beta \dot{\beta}\right)
\end{equation}

\noindent where the $\omega$ on the left side comes from the uniform rotational motion of $\mathbf{r}_0$.  After replacing $\dot{\beta}$ by $\omega\beta$ and subtracting on both sides $\omega$, can be rewritten into

\begin{equation}
    \frac{d\psi}{dt'} = \gamma^2\omega\left(\beta \sin\psi - 1\right)
\end{equation}

\noindent Again, this equation admits an analytical solution which reads

\begin{equation}
\begin{split}
    &\psi(t') =\; 2 \arctan\bigg\{\beta-\\
    &\frac{1}{\gamma}\tan\left[\frac{\gamma \omega (t'-t)}{2}+\arctan\left[ \gamma \left(\beta-\tan\frac{\phi}{2}\right) \right] \right] \bigg\}
\end{split}
\end{equation}

\noindent where $\phi$ now is the final value of $\psi(t')$ for $t'=t$. 
\begin{figure}
\centering
\includegraphics[width=6.5cm]{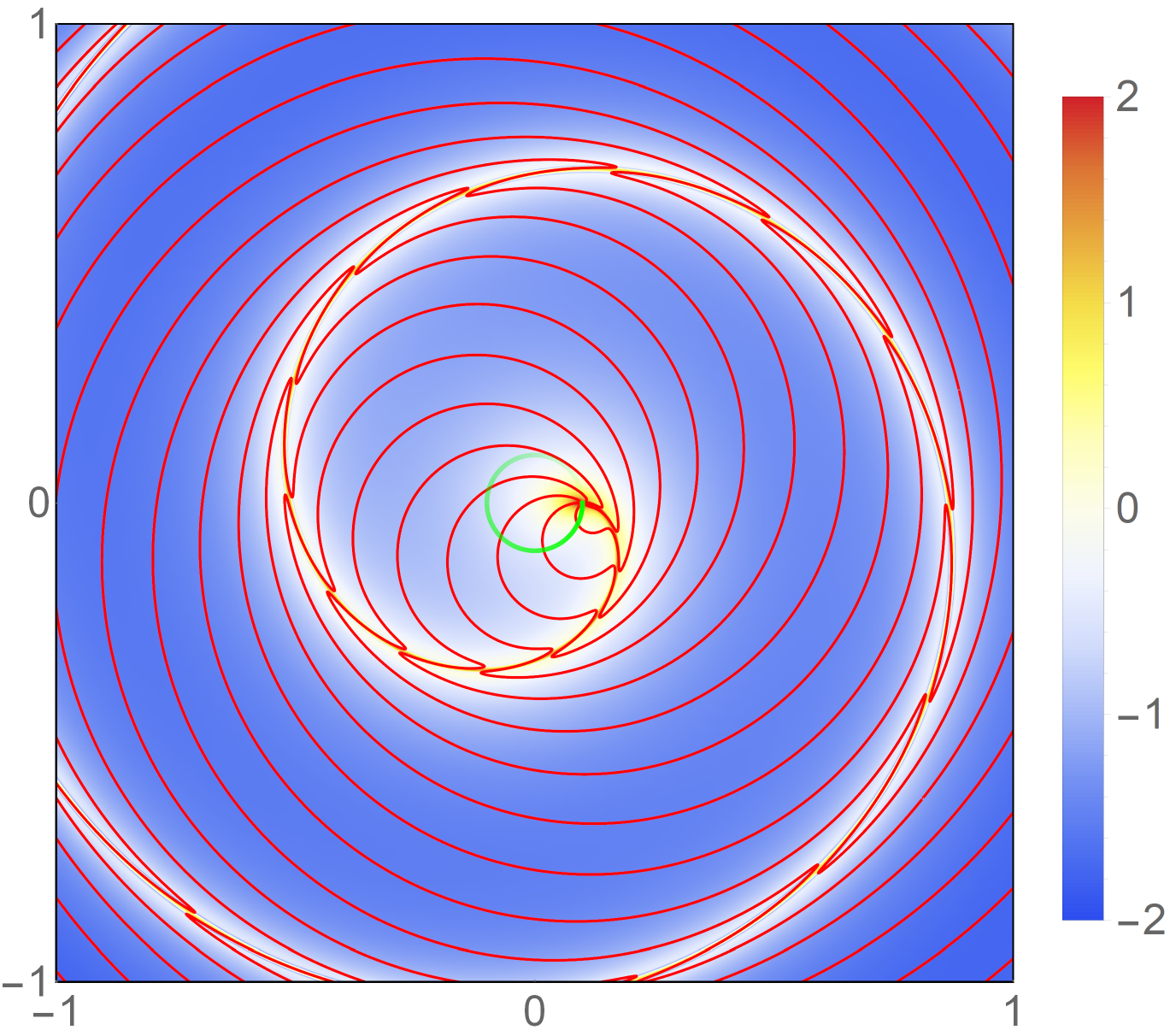}
\caption{Same as Fig.~\ref{fig:Fig6} but for a circularly moving point charge at 0.9 light speed.}
\label{fig:Fig7}
\end{figure}
Now, with the solution for $\psi(t')$ at hand, the unit vector $\hat{\boldsymbol{\lambda}}$ is given in Cartesian $(x,y)$-coordinates by 

\begin{equation}
\begin{split}
    \hat{\boldsymbol{\lambda}}(t') = \left\{\cos\left[\omega t' + \psi(t')\right], \sin\left[\omega t' + \psi(t')\right] \right\}
\end{split}
\end{equation}

\noindent which again solves the full problem. As a numerical example, let us first consider the case of charge moving with 0.6 light speed around a circle of $a=100$~m. Thus, we find for the angular frequency a value of $\omega = 0.6\;c/a \sim 1.8\cdot10^6~$s$^{-1}$, which corresponds to an oscillation period of $\sim3.5$~\textmu s. The resulting field line structure and electric field are presented in Fig.~\ref{fig:Fig6}. For comparison, we consider also a travel speed of $0.9\;c$, which corresponds to an angular frequency of $\omega = 0.9\;c/a \sim 1.8\cdot10^6~$s$^{-1}$, or an oscillation period of $\sim2.3$~\textmu s. The field line structure and electric field for this case are shown in Fig.~\ref{fig:Fig7}. Although both figures \ref{fig:Fig6} and \ref{fig:Fig7} look qualitatively similar, one can see how the field line structures develops a more pronounced shock-wave structure for velocities closer to the speed of light.

\section{Conclusion}

We have presented an elementary derivation of a differential equation, eq.~\eqref{eq:lambda}, the solution of which leads to a simple description of electric field lines for an arbitrarily moving charge. We have presented several analytical solution of this equation for a quite broad class of cases. Even if one cannot find analytical solutions to eq.~\eqref{eq:lambda}, its simplicity should make numerical integration straightforward. Thus, it provides a powerful tool for visualizing the electric field structure generated by a point charge moving along arbitrarily complex and relativistic trajectories.

\begin{acknowledgments}
We a grateful to the Deutsche Forschungsgemeinschaft (DFG) for financial support via projects A14 (SFB 937) and A05 (SFB 755).
\end{acknowledgments}


\begin{thebibliography}{22}%
\makeatletter
\providecommand \@ifxundefined [1]{%
 \@ifx{#1\undefined}
}%
\providecommand \@ifnum [1]{%
 \ifnum #1\expandafter \@firstoftwo
 \else \expandafter \@secondoftwo
 \fi
}%
\providecommand \@ifx [1]{%
 \ifx #1\expandafter \@firstoftwo
 \else \expandafter \@secondoftwo
 \fi
}%
\providecommand \natexlab [1]{#1}%
\providecommand \enquote  [1]{``#1''}%
\providecommand \bibnamefont  [1]{#1}%
\providecommand \bibfnamefont [1]{#1}%
\providecommand \citenamefont [1]{#1}%
\providecommand \href@noop [0]{\@secondoftwo}%
\providecommand \href [0]{\begingroup \@sanitize@url \@href}%
\providecommand \@href[1]{\@@startlink{#1}\@@href}%
\providecommand \@@href[1]{\endgroup#1\@@endlink}%
\providecommand \@sanitize@url [0]{\catcode `\\12\catcode `\$12\catcode
  `\&12\catcode `\#12\catcode `\^12\catcode `\_12\catcode `\%12\relax}%
\providecommand \@@startlink[1]{}%
\providecommand \@@endlink[0]{}%
\providecommand \url  [0]{\begingroup\@sanitize@url \@url }%
\providecommand \@url [1]{\endgroup\@href {#1}{\urlprefix }}%
\providecommand \urlprefix  [0]{URL }%
\providecommand \Eprint [0]{\href }%
\providecommand \doibase [0]{http://dx.doi.org/}%
\providecommand \selectlanguage [0]{\@gobble}%
\providecommand \bibinfo  [0]{\@secondoftwo}%
\providecommand \bibfield  [0]{\@secondoftwo}%
\providecommand \translation [1]{[#1]}%
\providecommand \BibitemOpen [0]{}%
\providecommand \bibitemStop [0]{}%
\providecommand \bibitemNoStop [0]{.\EOS\space}%
\providecommand \EOS [0]{\spacefactor3000\relax}%
\providecommand \BibitemShut  [1]{\csname bibitem#1\endcsname}%
\let\auto@bib@innerbib\@empty
%</preamble>
\bibitem [{\citenamefont {Landau}\ and\ \citenamefont
  {Lifshitz}(1971)}]{landau1971classical}%
  \BibitemOpen
  \bibfield  {author} {\bibinfo {author} {\bibfnamefont {L.}~\bibnamefont
  {Landau}}\ and\ \bibinfo {author} {\bibfnamefont {E.}~\bibnamefont
  {Lifshitz}},\ }\enquote {\bibinfo {title} {The classical theory of fields},}\
  \ (\bibinfo  {publisher} {Pergamon},\ \bibinfo {year} {1971})\ Chap.~\bibinfo
  {chapter} {3}, pp.\ \bibinfo {pages} {1--10}\BibitemShut {NoStop}%
\bibitem [{\citenamefont {Jackson}(1998)}]{jackson1971classical}%
  \BibitemOpen
  \bibfield  {author} {\bibinfo {author} {\bibfnamefont {J.}~\bibnamefont
  {Jackson}},\ }\enquote {\bibinfo {title} {Classical electrodynamics},}\ \
  (\bibinfo  {publisher} {John Wiley \& Sons},\ \bibinfo {year} {1998})\
  Chap.~\bibinfo {chapter} {14}, pp.\ \bibinfo {pages} {661--707}\BibitemShut
  {NoStop}%
\bibitem [{\citenamefont {Tsien}(1972)}]{tsien1972pictures}%
  \BibitemOpen
  \bibfield  {author} {\bibinfo {author} {\bibfnamefont {R.~Y.}\ \bibnamefont
  {Tsien}},\ }\href@noop {} {\bibfield  {journal} {\bibinfo  {journal}
  {American Journal of Physics}\ }\textbf {\bibinfo {volume} {40}},\ \bibinfo
  {pages} {46} (\bibinfo {year} {1972})}\BibitemShut {NoStop}%
\bibitem [{\citenamefont {Ohanian}(1980)}]{ohanian1980electromagnetic}%
  \BibitemOpen
  \bibfield  {author} {\bibinfo {author} {\bibfnamefont {H.~C.}\ \bibnamefont
  {Ohanian}},\ }\href@noop {} {\bibfield  {journal} {\bibinfo  {journal}
  {American Journal of Physics}\ }\textbf {\bibinfo {volume} {48}},\ \bibinfo
  {pages} {170} (\bibinfo {year} {1980})}\BibitemShut {NoStop}%
\bibitem [{\citenamefont {Jefimenko}(1994)}]{jefimenko1994direct}%
  \BibitemOpen
  \bibfield  {author} {\bibinfo {author} {\bibfnamefont {O.~D.}\ \bibnamefont
  {Jefimenko}},\ }\href@noop {} {\bibfield  {journal} {\bibinfo  {journal}
  {American Journal of Physics}\ }\textbf {\bibinfo {volume} {62}},\ \bibinfo
  {pages} {79} (\bibinfo {year} {1994})}\BibitemShut {NoStop}%
\bibitem [{\citenamefont {Bolotovskii}\ and\ \citenamefont
  {Serov}(1997)}]{bolotovskiui1997force}%
  \BibitemOpen
  \bibfield  {author} {\bibinfo {author} {\bibfnamefont {B.}~\bibnamefont
  {Bolotovskii}}\ and\ \bibinfo {author} {\bibfnamefont {A.}~\bibnamefont
  {Serov}},\ }\href@noop {} {\bibfield  {journal} {\bibinfo  {journal}
  {Physics-Uspekhi}\ }\textbf {\bibinfo {volume} {40}},\ \bibinfo {pages}
  {1055} (\bibinfo {year} {1997})}\BibitemShut {NoStop}%
\bibitem [{\citenamefont {Singal}(2011)}]{singal2011first}%
  \BibitemOpen
  \bibfield  {author} {\bibinfo {author} {\bibfnamefont {A.~K.}\ \bibnamefont
  {Singal}},\ }\href@noop {} {\bibfield  {journal} {\bibinfo  {journal}
  {American Journal of Physics}\ }\textbf {\bibinfo {volume} {79}},\ \bibinfo
  {pages} {1036} (\bibinfo {year} {2011})}\BibitemShut {NoStop}%
\bibitem [{\citenamefont {Datta}(2016)}]{datta2016b}%
  \BibitemOpen
  \bibfield  {author} {\bibinfo {author} {\bibfnamefont {S.}~\bibnamefont
  {Datta}},\ }\href@noop {} {\bibfield  {journal} {\bibinfo  {journal} {Physics
  Education}\ }\textbf {\bibinfo {volume} {32}},\ \bibinfo {pages} {2}
  (\bibinfo {year} {2016})}\BibitemShut {NoStop}%
\bibitem [{\citenamefont {Franklin}\ and\ \citenamefont
  {Ryder}(2019)}]{franklin2019electromagnetic}%
  \BibitemOpen
  \bibfield  {author} {\bibinfo {author} {\bibfnamefont {J.}~\bibnamefont
  {Franklin}}\ and\ \bibinfo {author} {\bibfnamefont {A.}~\bibnamefont
  {Ryder}},\ }\href@noop {} {\bibfield  {journal} {\bibinfo  {journal}
  {American Journal of Physics}\ }\textbf {\bibinfo {volume} {87}},\ \bibinfo
  {pages} {153} (\bibinfo {year} {2019})}\BibitemShut {NoStop}%
\bibitem [{mma()}]{mmaprogram}%
  \BibitemOpen
  \href@noop {} {}\bibinfo {howpublished}
  {https://www.dropbox.com/s/c6ofgx6aod93tva/\\ElectricFieldlines.nb?dl=0}\BibitemShut
  {NoStop}%
\bibitem [{mov()}]{movies}%
  \BibitemOpen
  \href@noop {} {}\bibinfo {howpublished}
  {https://www.joerg-enderlein.de/relativistic-charge}\BibitemShut {NoStop}%
\bibitem [{\citenamefont {Arutyunyan}(1985)}]{arutyunyan1985force}%
  \BibitemOpen
  \bibfield  {author} {\bibinfo {author} {\bibfnamefont {S.}~\bibnamefont
  {Arutyunyan}},\ }\href@noop {} {\bibfield  {journal} {\bibinfo  {journal}
  {Radiophysics and Quantum Electronics}\ }\textbf {\bibinfo {volume} {28}},\
  \bibinfo {pages} {619} (\bibinfo {year} {1985})}\BibitemShut {NoStop}%
\bibitem [{\citenamefont {Arutyunyan}(1986)}]{arutyunyan1986electromagnetic}%
  \BibitemOpen
  \bibfield  {author} {\bibinfo {author} {\bibfnamefont {S.}~\bibnamefont
  {Arutyunyan}},\ }\href@noop {} {\bibfield  {journal} {\bibinfo  {journal}
  {Physics-Uspekhi}\ }\textbf {\bibinfo {volume} {29}},\ \bibinfo {pages}
  {1053} (\bibinfo {year} {1986})}\BibitemShut {NoStop}%
\bibitem [{\citenamefont {Arutyunyan}(1992)}]{arutyunyan1992equivavlent}%
  \BibitemOpen
  \bibfield  {author} {\bibinfo {author} {\bibfnamefont {S.}~\bibnamefont
  {Arutyunyan}},\ }\href@noop {} {\bibfield  {journal} {\bibinfo  {journal}
  {Izvestiya Vysshikh Uchebnykh Zavedenii, Radiofizika}\ }\textbf {\bibinfo
  {volume} {35}},\ \bibinfo {pages} {313} (\bibinfo {year} {1992})}\BibitemShut
  {NoStop}%
\bibitem [{\citenamefont {Hopf}\ \emph {et~al.}(1976)\citenamefont {Hopf},
  \citenamefont {Meystre}, \citenamefont {Scully},\ and\ \citenamefont
  {Louisell}}]{hopf1976classical}%
  \BibitemOpen
  \bibfield  {author} {\bibinfo {author} {\bibfnamefont {F.}~\bibnamefont
  {Hopf}}, \bibinfo {author} {\bibfnamefont {P.}~\bibnamefont {Meystre}},
  \bibinfo {author} {\bibfnamefont {M.}~\bibnamefont {Scully}}, \ and\ \bibinfo
  {author} {\bibfnamefont {W.}~\bibnamefont {Louisell}},\ }\href@noop {}
  {\bibfield  {journal} {\bibinfo  {journal} {Physical Review Letters}\
  }\textbf {\bibinfo {volume} {37}},\ \bibinfo {pages} {1215} (\bibinfo {year}
  {1976})}\BibitemShut {NoStop}%
\bibitem [{\citenamefont {Colson}(1977)}]{colson1977one}%
  \BibitemOpen
  \bibfield  {author} {\bibinfo {author} {\bibfnamefont {W.}~\bibnamefont
  {Colson}},\ }\href@noop {} {\bibfield  {journal} {\bibinfo  {journal}
  {Physics Letters A}\ }\textbf {\bibinfo {volume} {64}},\ \bibinfo {pages}
  {190} (\bibinfo {year} {1977})}\BibitemShut {NoStop}%
\bibitem [{\citenamefont {Brau}(1996)}]{brau1996undulator}%
  \BibitemOpen
  \bibfield  {author} {\bibinfo {author} {\bibfnamefont {C.~A.}\ \bibnamefont
  {Brau}},\ }\href@noop {} {\bibfield  {journal} {\bibinfo  {journal} {American
  Journal of Physics}\ }\textbf {\bibinfo {volume} {64}},\ \bibinfo {pages}
  {662} (\bibinfo {year} {1996})}\BibitemShut {NoStop}%
\bibitem [{\citenamefont {Winick}\ and\ \citenamefont
  {Bienenstock}(1978)}]{winick1978synchrotron}%
  \BibitemOpen
  \bibfield  {author} {\bibinfo {author} {\bibfnamefont {H.}~\bibnamefont
  {Winick}}\ and\ \bibinfo {author} {\bibfnamefont {A.}~\bibnamefont
  {Bienenstock}},\ }\href@noop {} {\bibfield  {journal} {\bibinfo  {journal}
  {Annual Review of Nuclear and Particle Science}\ }\textbf {\bibinfo {volume}
  {28}},\ \bibinfo {pages} {33} (\bibinfo {year} {1978})}\BibitemShut {NoStop}%
\bibitem [{\citenamefont {Hofmann}(2004)}]{hofmann2004physics}%
  \BibitemOpen
  \bibfield  {author} {\bibinfo {author} {\bibfnamefont {A.}~\bibnamefont
  {Hofmann}},\ }\href@noop {} {\emph {\bibinfo {title} {The physics of
  synchrotron radiation}}},\ Vol.~\bibinfo {volume} {20}\ (\bibinfo
  {publisher} {Cambridge University Press},\ \bibinfo {year}
  {2004})\BibitemShut {NoStop}%
\bibitem [{\citenamefont {Hannay}\ and\ \citenamefont
  {Jeffrey}(2005)}]{hannay2005electric}%
  \BibitemOpen
  \bibfield  {author} {\bibinfo {author} {\bibfnamefont {J.}~\bibnamefont
  {Hannay}}\ and\ \bibinfo {author} {\bibfnamefont {M.}~\bibnamefont
  {Jeffrey}},\ }\href@noop {} {\bibfield  {journal} {\bibinfo  {journal}
  {Proceedings of the Royal Society A: Mathematical, Physical and Engineering
  Sciences}\ }\textbf {\bibinfo {volume} {461}},\ \bibinfo {pages} {3599}
  (\bibinfo {year} {2005})}\BibitemShut {NoStop}%
\bibitem [{\citenamefont {Novokhatski}(2011)}]{novokhatski2011field}%
  \BibitemOpen
  \bibfield  {author} {\bibinfo {author} {\bibfnamefont {A.}~\bibnamefont
  {Novokhatski}},\ }\href@noop {} {\bibfield  {journal} {\bibinfo  {journal}
  {Physical Review Special Topics-Accelerators and Beams}\ }\textbf {\bibinfo
  {volume} {14}},\ \bibinfo {pages} {060707} (\bibinfo {year}
  {2011})}\BibitemShut {NoStop}%
\bibitem [{\citenamefont {Patterson}(2011)}]{patterson2011simplified}%
  \BibitemOpen
  \bibfield  {author} {\bibinfo {author} {\bibfnamefont {B.}~\bibnamefont
  {Patterson}},\ }\href@noop {} {\bibfield  {journal} {\bibinfo  {journal}
  {American Journal of Physics}\ }\textbf {\bibinfo {volume} {79}},\ \bibinfo
  {pages} {1046} (\bibinfo {year} {2011})}\BibitemShut {NoStop}%
\end{thebibliography}
\end{document}